\documentclass[a4paper,10pt]{llncs}

\usepackage{graphicx}

\title{Extending Probabilistic Data Fusion Using Sliding Windows}
\author{David Lillis\inst{1} \and Fergus Toolan\inst{2} \and Rem Collier\inst{1} \and John Dunnion\inst{1}}

\institute{School of Computer Science and Informatics\\ University College Dublin \\ \email{\{david.lillis, rem.collier, john.dunnion\}@ucd.ie}
\and Department of Computing Science\\ Griffith College Dublin \\ \email{fergus.toolan@gcd.ie}}

\begin{document}

\bibliographystyle{splncs}

\maketitle

\begin{abstract}
Recent developments in the field of data fusion have seen a focus on techniques that use training queries to estimate the probability that various documents are relevant to a given query and use that information to assign scores to those documents on which they are subsequently ranked. This paper introduces SlideFuse, which builds on these techniques, introducing a sliding window in order to compensate for situations where little relevance information is available to aid in the estimation of probabilities.

SlideFuse is shown to perform favourably in comparison with CombMNZ, ProbFuse and SegFuse. CombMNZ is the standard baseline technique against which data fusion algorithms are compared whereas ProbFuse and SegFuse represent the state-of-the-art for probabilistic data fusion methods.
\end{abstract}

\section{Introduction} \label{introduction}

The aim of any Information Retrieval (IR) system is the identification of documents that best satisfy a user's information need, typically expressed in terms of a textual query. Traditional approaches to IR employ algorithms responsible for analysing the contents of the documents themselves in order to return those that most closely relate to the query provided.

More recently, there is a growing body of research focused on combining the output of several such systems with the aim of creating a single set of results that will have greater relevance than the output of any individual system \cite{Bartell1994,Beitzel2004,Vogt1999}. Algorithms to perform this type of combination vary according to the situations in which they are intended to be used. This paper concentrates on the ``data fusion'' family of algorithms, which are intended for use in cases where each input system has access to the same document collections \cite{Aslam2000}. This is distinct from ``collection fusion'' \cite{Voorhees1994}, where the document collections are disjoint, or cases where only partial overlap exists between collections.

The principal difference between these situations is that data fusion algorithms may consider the presence of a document in multiple result sets as evidence of relevance, since a document's absence in a result set can only be as a result of it not being considered relevant by the corresponding input system. In contrast, where the overlap between document collections is not complete, the absence of a document from a result set may merely reflect its absence from the underlying document collection and so is not necessarily a reliable indication that the document has been considered to be nonrelevant.

This paper introduces SlideFuse, a novel probabilistic data fusion algorithm that uses the past performance of its underlying input systems as an indication of the probability that certain documents will be relevant to future queries. This assumption has been previously demonstrated to achieve favourable results \cite{Lillis2006b,Shokouhi2006}. It is robust in the face of incomplete training data by utilising information about a document's neighbours as evidence of its likelihood of relevance. It does this while avoiding some of the shortfalls of existing probabilistic methods.

Section~\ref{background} gives a brief outline of previous research in the area of data fusion. In Section~\ref{slideintro}, we present an overview of how SlideFuse operates, followed by a formal definition of the algorithm in Section~\ref{slidedefn}. Section~\ref{setup} details the setup of the experiments that were run to evaluate the SlideFuse algorithm, the results of which are presented in Section~\ref{results}. This includes a comparison with the CombMNZ algorithm, which is a standard baseline frequently used in data fusion research, as well as ProbFuse and SegFuse, two recent probabilistic data fusion techniques. Finally, conclusions and future work are discussed in Section~\ref{conclusion}.

\section{Data Fusion} \label{background}

Traditionally, data fusion techniques fall into two broad categories: score-based fusion and rank-based fusion. Score-based techniques make use of the scores each input system uses to rank the documents in its result set. This typically necessitates the use of some form of score normalisation \cite{Fox1994}, in order to ensure that the results cannot be skewed by the use of different methods of allocating scores (e.g. one input system may score documents on a scale of 0-100 whereas another may use a scale of 0-1).

A popular approach to score-based fusion is the use of a Linear Combination \cite{Bartell1994,Callan1995,Vogt1999}. Here, weights are attached to each input system, which are multipled by the ranking scores assigned each document. The final score for each document is the sum of these. Normalised scores have also been used in this context \cite{Si2002}.

An important suite of data fusion techniques based on normalised scores was proposed in \cite{Fox1994}. Of these, CombMNZ has become the standard data fusion technique against which new algorithms are compared \cite{Beitzel2004,Montague2002}. Here, the final score assigned to each document is the sum of the normalised scores it is given in each input result set, multiplied by the number of input systems that returned it. Significant work was carried out by Lee to demonstrate CombMNZ's effectiveness \cite{Lee1997}.

Interleaving is perhaps the simplest rank-based fusion technique \cite{Voorhees1994}. This involves removing the top document from each input result set in turn and adding it to the fused set to be returned. Weighted variations on this have also been proposed so as to benefit input systems that have achieved superior performance in the past \cite{Voorhees1995}. Two voting-based techniques based on document ranks were proposed by Aslam and Montague \cite{Montague2002,Aslam2001}. These used the analogy of the input systems representing few electors and the documents representing many candidates to be ranked.

An algorithm making use of the textual contents of the documents was presented in \cite{Craswell1999,Lawrence1998}. Another relies on the input systems providing metadata relating to the documents they return, which can be used in the fusion process \cite{Gravano1997}.

In recent times, a variation of rank-based fusion has emerged, whereby result sets are divided into segments and documents are assigned a score based on the segments in which they appear, rather than their exact rank within the result set. The ProbFuse algorithm  \cite{Lillis2006b,Lillis2006c} divides each result set into equal length segments and uses training data to estimate the probability that a document returned in a particular segment by a particular input system is relevant. This is done by calculating the proportion of documents returned in each segment by each input system that are relevant to the training queries, compared to nonrelevant documents.

A similar approach is taken with SegFuse \cite{Shokouhi2006}, with the major exception being that the segments are not of equal length, but rather increase in size exponentially later in the result set. As relevant documents are most likely to occur in the early part of a result set, maintaining small segment sizes in early positions advantages these early documents, as they are less likely to be grouped with less relevant documents occurring later on. SegFuse also takes normalised scores into account.

\section{SlideFuse: Introduction} \label{slideintro}

Existing segment-based data fusion techniques ProbFuse and SegFuse use the probability that a document is relevant to assign a score on which it is eventually ranked in the final result set. This probability is estimated by analysing the results of a number of training queries for which relevance judgments are available. Relevance judgments are typically included with IR test collections, and specify which documents in the collection have been judged to be relevant, or nonrelevant, to test queries. However, with large document collections these judgments tend to be incomplete, meaning that only relatively few documents have been judged for each query, leaving the majority unjudged. This incompleteness causes difficulty in analysing training data, as there may be positions in result sets in which a document that is known to be relevant is never returned, though this does not necessarily entail that a relevant document is never located at that rank. 

\begin{figure}[!hb]
\begin{center}
\includegraphics[scale=0.25]{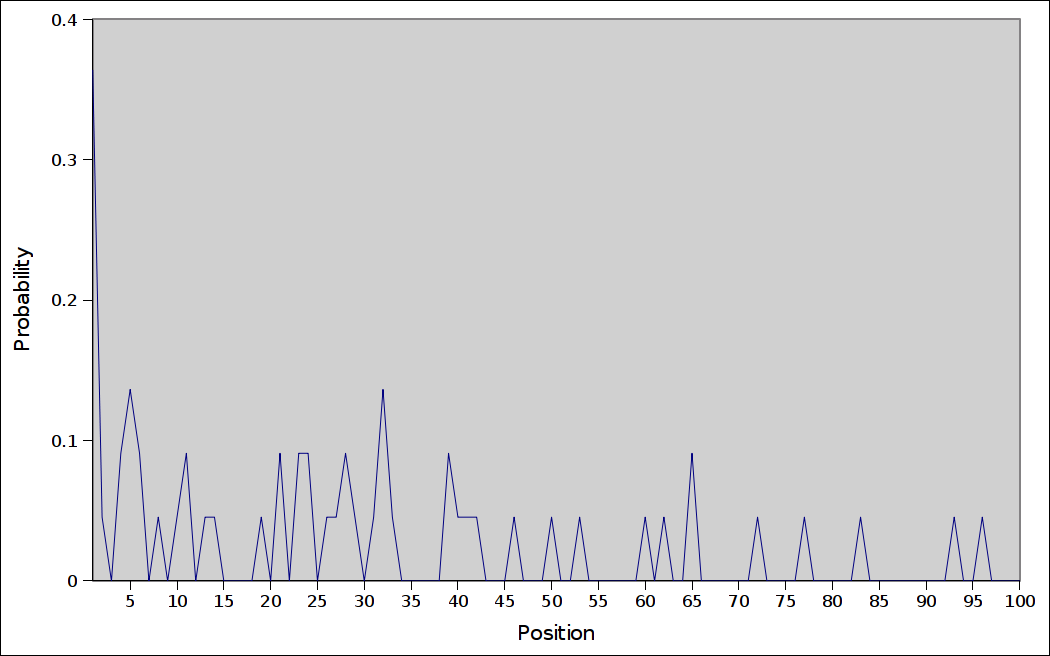}

\caption{Probability Distribution using Individual Positions}\label{probposition}
\end{center}
\end{figure}

For this reason, calculating probabilities at the individual rank level results in an extremely jagged probability distribution. For instance, with the Web Track from the TREC-2004 conference (which is the document collection used in the experiments presented in Section~\ref{setup}), calculating the probability for each position results in the graph presented in Figure~\ref{probposition}. In that figure, the probability value used in each position is the number of relevant documents returned in that postion over all the training queries, divided by the total number of training queries that returned a document in that position (i.e. a result set of only 100 documents in length will not have returned a document in position 101).

One motivation behind segmenting result sets is to counter this effect, by not estimating the probability of relevance of a document returned at a particular rank solely based on documents returned at that exact rank for the training queries. Instead, relevant documents returned at other positions within the same segment are also taken into account, so smoothing the distribution of probability scores.

\begin{figure}[!hb]
\begin{center}
\includegraphics[scale=0.25]{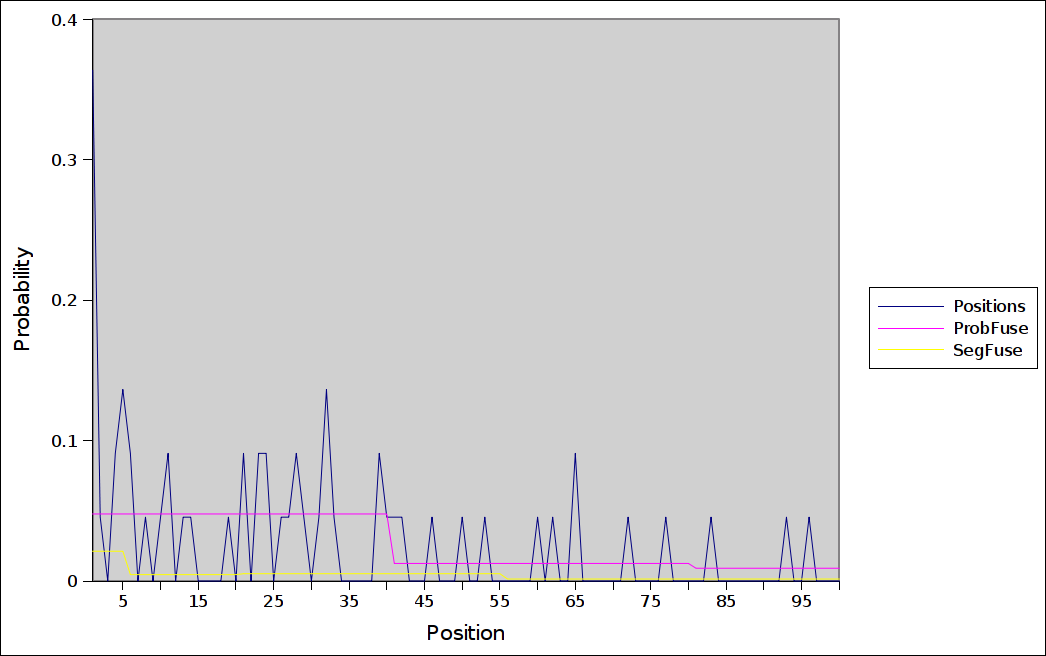}

\caption{Probability Distribution using ProbFuse and SegFuse}\label{probseg}
\end{center}
\begin{center}
\includegraphics[scale=0.25]{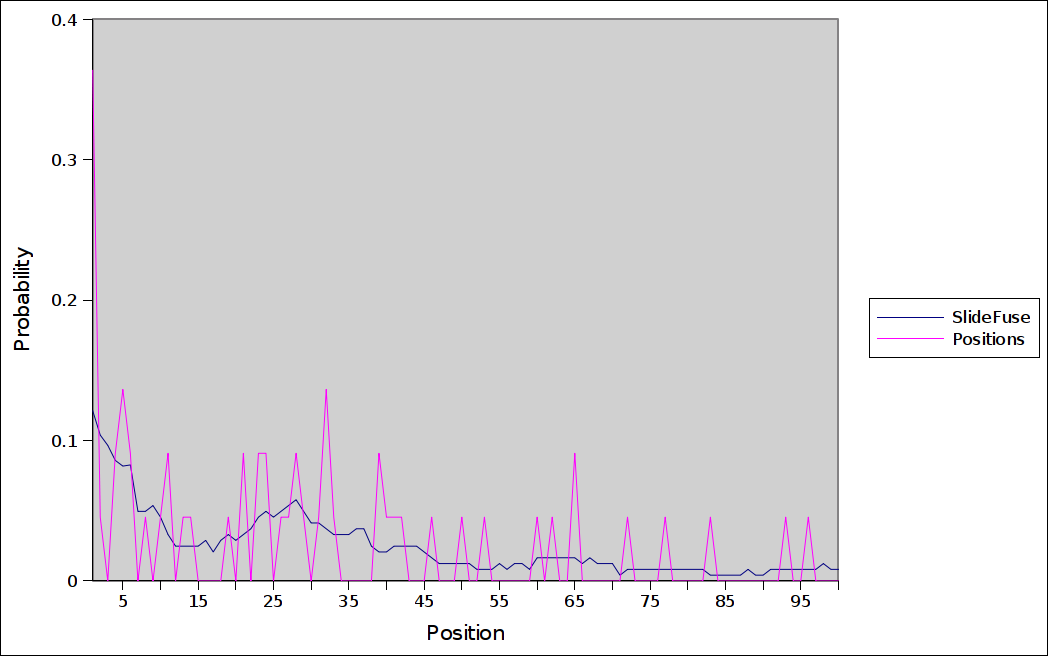}

\caption{Probability Distribution using SlideFuse}\label{slide}
\end{center}
\end{figure}

One consequence of this approach is that it is possible for a significant drop in probability score to occur at the boundary between segments. This effect is illustrated in Figure~\ref{probseg}. For example, in a result set divided into segments of 40 documents each, the probability associated with the document returned in position 40 is likely to be much higher than that of the document returned in position 41. This is because the probability for the segment containing position 40 is calculated using positions 1 through 40, whereas the segment containing position 41 ranges from position 41 to position 80. As the former encompasses documents much higher in the result set (that are more likely to be relevant than documents further down the result set), position 40 is given an artificial advantage over position 41. This is easily demonstrated by plotting a graph of probability score against position. Unlike ProbFuse, SegFuse changes the size of each segment in different areas of the result set, with the smallest segments being at the beginning. This has the effect of reducing the distance between such segment boundaries at the beginning of the result set, where relevant documents are most likely to appear and consequently reducing the occurrence of sudden changes in probability scores.

In order to address the problem of the sudden drops in the probability scores associated with segmentation, and the problem of incomplete relevance judgments, we introduce the concept of a window surrounding each rank, where the probability assigned to that rank is based on the proportion of relevant documents located in its surrounding window during the training phase. For example, if we define the size of the sliding window to extend to 5 documents on either side of the relevant position, the window for rank 40 will extend from position 35 to position 45 inclusive. Similarly, the sliding window for position 41 extends from rank 36 to rank 46. With this approach, the problem of the location of segment boundaries is eliminated, as it is the closest neighbouring positions that are always taken into account. The effect on probability distribution is shown in Figure~\ref{slide}. As a special case, SlideFuse ensures that a sliding window cannot extend beyond the boundaries of the result set.

The use of training data entails that the scores that are ultimately assigned to each document are based on the past performance of each input system, thus encompassing an implied weighting system wereby documents returned by input systems with a prior record of greater effectiveness will receive a higher score. A formal mathematical description is presented in Section~\ref{slidedefn}.

\section{SlideFuse: Description} \label{slidedefn}

In common with other probabilistic data fusion techniques, SlideFuse requires both a training phase and a fusion phase. In the training phase, relevance information is gleaned from result sets returned in response to training queries for which relevance judgments are available. Later, this training data is used to fuse result sets produced by the same input systems relating to other queries.

\subsection{Training Phase: Rank Probability}
The training phase consists of estimating for each input system the probability that a document returned in any given rank in that system's result set is relevant.

Formally, $P(d_p|s)$, the probability that a document $d$ returned in position $p$ of a result set is relevant, given that is has been returned by input system $s$ is given by

\begin{equation}
P(d_p|s) = \frac{\sum_{q \in Q_p} R_{d_p,q}}{Q_p}
\end{equation}

where $Q_p$ is the set of all training queries for which at least $p$ documents were returned by the input system and $R_{d_p,q}$ is the relevance of the document $d_p$ to query $q$ (1 if the document is relevant, 0 if not). This is calculated for each input system to be used in the fusion phase.

\subsection{Fusion Phase: Window Boundaries}

As noted in Section~\ref{slideintro}, using the probability at each rank leads to inconsistent results on document collections with incomplete relevance judgments, due to the high number of documents in each result set that have not been judged relevant (and are therefore assumed not to be relevant). In order to achieve more useful probability values, we construct a window around each position, so as to make use of relevance information about near neighbours when assigning probabilities to individual ranks.

The start and end points ($a$ and $b$ respectively) of the sliding window surrounding each result set position $p$ are given by

\begin{equation}
 a=\left\{
\begin {array}{lr}
p-w& \hspace{0.5cm} p-w>=0\\
\noalign{\medskip}
0&p-w<0
\end {array}
\right.
\end{equation}

\begin{equation}
 b=\left\{
\begin {array}{lr}
p+w&p+w<N\\
\noalign{\medskip}
N-1& \hspace{0.5cm} p+w>=N
\end {array}
\right.
\end{equation}

where $w$ is a parameter that indicates how many positions on either side of $p$ should be included in the window and $N$ is the total number of documents in the result set. In effect, the above definitions of $a$ and $b$ ensure that the window cannot begin before the first document in the result set and also cannot extend beyond the last document.

\subsection{Fusion Phase: Assigning Probabilities to Windows}

Once the window boundaries have been set around each position of each of the result sets that are to be fused, the next stage in the fusion process is to assign a probability score to each position based on those positions contained in the window surrounding it.

$P(d_{p,w}|s)$, the probability of relevance of document $d$ in position $p$ using a window size of $w$ documents either side of $p$, given that it has been returned by input system $s$ is given by

\begin{equation}
 P(d_{p,w}|s) = \frac{\sum_{i=a}^{b} P(d_i|s)}{b-a+1}
\end{equation}

The use of the sliding window results in a smoother decrease in the probabilities later in the result set, when compared with using probabilities based on data available at each position alone.

\subsection{Fusion Phase: Ranking Score}

Once the above stages have been completed, the final step is to assign a score to each document. $R_d$, the final ranking score given to document $d$ is given by

\begin{equation}
R_d = \sum_{s \in S} P(d_{p,w}|s)
\end{equation}

where $S$ is the set of all input systems used and $p$ is the position in which document $d$ was returned by input system $s$. Using the sum of the probability scores makes use of the ``Chorus Effect'', which argues that multiple input systems agreeing on the relevance of a document is evidence that the document in question is actually relevant \cite{Vogt1999}. The ``Skimming Effect'' is also important in the context of data fusion \cite{Vogt1999}. This states that since relevant documents are most likely to be located in early positions in a result set, weighting highly-ranked documents heavily is beneficial when performing fusion. Although there is no explicit consideration of this effect made in the definition of SlideFuse, the probability distribution in Figure~\ref{slide} shows that this increased likelihood of relevance in early positions automatically benefits these highly-ranked documents.

\section{Experiment Setup} \label{setup}

The document collection used for evaluation is the Web Track from the TREC-2004 conference \cite{Craswell2004}. A feature of this document collection is that the relevance judgments are extremely incomplete. The available data includes 74 topfiles (each containing result sets produced by a single input system in response to each of 225 queries). A number of measures were taken in order to reduce the possibility of any bias being introduced by either the selection of input systems or the ordering of the queries.

Five runs of the experiment were performed. For each run, six topfiles were selected and the result sets from those topfiles were fused using SlideFuse, ProbFuse, SegFuse and CombMNZ. No topfile was used in more than one experimental run, the result of which being that of the 64 topfiles available, 30 were used for the purposes of this experiment. So as to eliminate the possibility of the ordering of the queries introducing any sort of bias, each run was performed five times, with the queries being shuffled each time. After shuffling, the first 10\% of queries were used for the purposes of training SlideFuse, ProbFuse and SegFuse. As the CombMNZ algorithm does not require a training phase, these training queries were ignored for that technique. The evaluation results presented below for each run are the average evaluation results from all of the various query orderings.

When running ProbFuse, each result set was divided into 25 segments, as in \cite{Lillis2006b}. For the purposes of SlideFuse, the value of the $w$ parameter was set to 5 (i.e. 5 documents on both sides of each position were included in the window). It is desirable to use a small value for $w$, in order that the probabilities at each position are only influenced by positions that are close by. However, initial experiments showed that using windows that are too small failed to fully address the problem outlined in Section~\ref{slideintro}, as there were still positions for which probabilities could not be calculated due to a lack of available relevance judgments.

When performing the evaluation of the four data fusion techniques, three evaluation measures were used: \emph{Mean Average Precision (MAP)} is the mean of the precision scores obtained after each relevant document has been retrieved. Relevant documents that are not included in the result set are given a precision of zero. MAP assumes that documents that have not been judged are nonrelevant. The \emph{bpref} measure evaluates the relative position of relevant and nonrelevant documents, ignoring documents that are unjudged. It was proposed by Buckley and Voorhees to cater for situations where relevance judgments are incomplete \cite{Buckley2004}. \emph{P10} measures the precision after 10 documents have been returned. Research has demonstrated that the vast majority of users of IR systems only examine the top 10 documents presented to them \cite{Silverstein1998}. Thus, the P10 measure places emphasis on documents returned in those positions where they are likely to be of use to the user.

Table~\ref{variation} illustrates the results of initial experiments aimed at choosing an appropriate training set size. Fusion was performed using each algorithm, with the training set sizes set to 10\%, 20\%, 30\%, 40\% and 50\% of available queries in turn. The performance of each algorithm was evaluated for each training set size using MAP. The Coefficient of Variation relating to these scores was then calculated for each algorithm for each run. This reflects the degree to which fusion performance is affected by changing the training set size. As Table~\ref{variation} illustrates, altering the number of training queries did not have any substantial effect on the performance of any of the fusion algorithms. Similar results were obtained for the bpref and P10 evaluation measures. Using only 10\% of the available queries for training thus reduces the amount of training data without adversely affecting performance.

\begin{table}
\begin{center}
\caption{Coefficient of Variation for MAP scores using training set sizes of 10\%, 20\%, 30\%, 40\%, 50\%}
\begin{tabular}{l |@{\hspace{10px}} l @{\hspace{10px}} |@{\hspace{10px}} l @{\hspace{10px}} | @{\hspace{10px}}l@{\hspace{10px}} | @{\hspace{10px}}l@{\hspace{10px}}} \label{variation}
	 	&	CombMNZ	 & ProbFuse &	SegFuse &	SlideFuse \\
\hline
first &		0.0033 &	0.0131 &	0.0056 &	0.0056 \\
second &	0.0056 &	0.1188 &	0.0581 &	0.0104 \\
third &		0.0380 &	0.0168 &	0.0120 &	0.0103 \\
fourth &	0.0034 &	0.0143 &	0.0111 &	0.0019 \\
fifth &		0.0246 &	0.0229 &	0.0491 &	0.0179 \\
\end{tabular}
\end{center}
\end{table}

\section{Analysis of Results} \label{results}

The results of comparing SlideFuse with CombMNZ, ProbFuse and SegFuse are shown in Tables~\ref{MAPresults}, \ref{BPREFresults} and \ref{P10results}. Each table presents the results from each of the five runs, along with the average result for each fusion technique. The ``vs. Best'' column displays the percentage difference between SlideFuse and the best of the other techniques (which is highlighted in bold in each case). The average in that column is the percentage difference between the average SlideFuse score and the best average score amongst the other algorithms. Values marked with ``*'' are statistically significant for a significance level of 5\%, using a paired t-test. Entries marked with ``**'' are significant for a significance level of 1\%.


\begin{table*}[!htb]
\caption{TREC-2004 performance of five individual runs evaluated with MAP}
\begin{center}
\begin{tabular}{l |@{\hspace{10px}} l @{\hspace{10px}} |@{\hspace{10px}} l @{\hspace{10px}} | @{\hspace{10px}}l@{\hspace{10px}} | @{\hspace{10px}}l@{\hspace{10px}} | @{\hspace{10px}}l}  \label{MAPresults}

	& CombMNZ	& ProbFuse	& SegFuse	& SlideFuse	& vs. Best \\
\hline
first	& 0.1598	& \textbf{0.4045}	& 0.1789	& 0.4977	& 23.05\% **\\
second	& 0.0783	& \textbf{0.2809}	& 0.1493	& 0.4905	& 74.58\% **\\
third	& 0.0426	& 0.2454	& \textbf{0.4946}	& 0.5103	& 3.17\% **\\
fourth	& 0.2454	& 0.2505	& \textbf{0.4995}	& 0.5025	& 0.61\% \\
fifth	& 0.1334	& 0.2892	& \textbf{0.3348}	& 0.3849	& 14.98\% ** \\
\hline
average	& 0.1319	& 0.2941	& \textbf{0.3314}	& 0.4772	& 43.99\% \\

\end{tabular}
\end{center}
\end{table*}

\begin{table*}[!htb]
\caption{TREC-2004 performance of five individual runs evaluated with bpref}
\begin{center}
\begin{tabular}{l |@{\hspace{10px}} l @{\hspace{10px}} |@{\hspace{10px}} l @{\hspace{10px}} | @{\hspace{10px}}l@{\hspace{10px}} | @{\hspace{10px}}l@{\hspace{10px}} | @{\hspace{10px}}l} \label{BPREFresults}

	& CombMNZ	& ProbFuse	& SegFuse	& SlideFuse	& vs. Best \\
\hline
first	& 0.2176		& \textbf{0.2997}	& 0.2547		& 0.4009	& 33.75\% **\\
second	& 0.3155		& 0.1877		& \textbf{0.3529}	& 0.4085	& 15.75\% **\\
third	& 0.1665		& 0.1281		& \textbf{0.4228}	& 0.4331	& 2.44\% *\\
fourth	& 0.4015		& 0.1375		& \textbf{0.4155}	& 0.4131	& -0.58\% \\
fifth	& 0.1945		& 0.1968		& \textbf{0.2971}	& 0.2996	& 0.83\% \\
\hline
average	& 0.2591		& 0.1900		& \textbf{0.3486}	& 0.3910	& 12.17\% \\

\end{tabular}
\end{center}

\caption{TREC-2004 performance of five individual runs evaluated with P10}
\begin{center}
\begin{tabular}{l |@{\hspace{10px}} l @{\hspace{10px}} |@{\hspace{10px}} l @{\hspace{10px}} | @{\hspace{10px}}l@{\hspace{10px}} | @{\hspace{10px}}l@{\hspace{10px}} | @{\hspace{10px}}l}  \label{P10results}

	& CombMNZ	& ProbFuse	& SegFuse	& SlideFuse	& vs. Best \\
\hline
first   & 0.1123        & \textbf{0.1344}       	& 0.1195        & 0.1413        & 5.15\% **\\
second  & 0.0349   	& \textbf{0.1023}       	& 0.0800       	& 0.1436       	& 40.39\% **\\
third   & 0.0257        & 0.1164       		& \textbf{0.1401}      	& 0.1445       	& 3.15\% **\\
fourth  & 0.1101       	& 0.1124       		& \textbf{0.1381}       & 0.1408        & 1.96\% *\\
fifth   & 0.0561       	& 0.1070		& \textbf{0.1113}      	& 0.1189        & 6.83\% **\\
\hline
average & 0.0678        & 0.1145       		& \textbf{0.1178}      	& 0.1378      	& 16.99\% \\

\end{tabular}
\end{center}
\end{table*}

Of the baseline techniques, ProbFuse performs best on the ``first'' and ``second'' runs, with SegFuse achieving superior performance on the others, with one exception in the bpref data. Overall, SlideFuse achieves the highest evaluation scores on average for all evaluation measures, with the single exception of the bpref score for the ``fourth'' run where the difference is 0.58\%, although this difference is not significant. Tests show that the performance improvements are statistically significant in most cases, with the exceptions being the ``fourth'' run when evaluated using MAP and the ``fifth'' run when evaluated with bpref.

Additionally, SlideFuse outperforms the best other technique in all runs using all three evaluation measures. When compared on an overall basis against any individual technique, the improvement is over 12\% in all cases, and is above 40\% when measured using MAP.

\section{Conclusions and Future Work} \label{conclusion}

This paper describes SlideFuse, a probabilistic data fusion algorithm that addresses some of the limitations of existing segment-based probabilistic techniques. On experiments using the TREC-2004 Web Track dataset, SlideFuse was shown to outperform the CombMNZ, ProbFuse and SegFuse data fusion techniques when evaluated using MAP, bpref and P10. Despite the fact that the training data available for the dataset is incomplete, SlideFuse was still capable of outperforming two algorithms that use the same training data (ProbFuse and SegFuse) and one that does not rely on training data (CombMNZ).

This was achieved by using a sliding window to use the probable relevance of a document's neighbours to estimate the probability that a document itself is relevant.

At present, SlideFuse assumes that each result set returned by an input system is of the same quality, as the probabilities used for fusion will be same in each case. In the future, we aim to investigate methods of weighting a particular result set according to its quality. This could possibly involve the use of the scores assigned to each document as a measure of an input system's confidence in its own results. Another approach to weighting would be to introduce weights within the sliding windows themselves, so as to place more emphasis on those documents that are closest to the rank around which the window is centred. Finally, a minor drawback of SlideFuse is that documents returned in positions beyond the length of the training sets will not be taken into account when fusing. We aim to address this situation in a more satisfactory fashion.

\bibliography{jabref}

\end{document}